\documentclass[10pt,a4paper,superscriptaddress,floatfix,twocolumn]{revtex4-1}

\usepackage{graphicx}
\usepackage{mathrsfs}
\usepackage{amsmath}
\usepackage[draft]{hyperref}

\begin{document}
\title{Hybrid optical and electronic laser locking using spectral hole burning}
\author{Warrick G. Farr, Jian Wei Tay, Patrick M. Ledingham, Dmitry Korystov, Jevon J. Longdell$^*$}
\address{Jack Dodd Centre, Department of Physics, University of Otago, 730 Cumberland Street, Dunedin, New Zealand \\
$^*$Corresponding author: jevon@physics.otago.ac.nz}

\begin{abstract}
 We report on a narrow linewidth laser diode system that is stabilized using both optical and electronic feedback to a spectral hole in cryogenic Tm:YAG. The laser system exhibits very low phase noise. The spectrum of the beat signal between two lasers, over millisecond timescales, is either Fourier limited or limited by the -111dBc/Hz noise floor. The resulting laser is well suited to quantum optics and sensing applications involving rare earth ion dopants.
\end{abstract}

\maketitle

Rare earth ion dopants in cryogenic crystals provide unique optical capabilities. They have very narrow optical homogeneous linewidths \cite{equa,75hz} while at the same time having much larger (multi GHz) inhomogeneous linewidths. In some rare earth systems long lived metastable states are available which leads to very effective optical pumping. The fine control of the lineshape of ensembles has lead to
recent quantum computing \cite{longdell} and quantum memory \cite{morgan,reid} demonstrations, and also the optical detection of ultrasound \cite{li08,tay10}.

Diode lasers provide low cost, convenient sources for use in many of these experiments. However the large amount of phase noise generally exhibited by diode lasers has limited their use, especially in ultrasound and continuous variable quantum optics applications. This phase noise is associated with the broad Schawlow-Townes linewidths exhibited by bare diode lasers, due to their short length and the low Q-factor of their resonators. They are also limited by the extreme sensitivity of their output frequency \cite{libbhall} to driving current.

There are two methods generally employed for reducing this noise level. One is to use electronic feedback to slave the laser to a more stable frequency reference \cite{pdh}. The other is to feedback some of the output light back into the laser diode after it has been through a dispersive optical network. This dispersion is in the sense that the phase shift varies with frequency. \cite{from_nist_paper}.

The narrow spectral holes possible in rare earth ion doped crystals have been used in a number of investigations as frequency references for Pound-Drever-Hall (PDH) electronic laser stabilization
\cite{bottger,pryde, julsgaard}.  Electronic stabilization techniques can no longer provide significant noise reduction at high frequencies because propagation delays and causality limit the servo bandwidths. This is not the case for optical feedback. Here as well as using PDH locking to lock the laser to a spectral hole we feed some of the light transmitted through the hole back into the laser diode. Spectral holes are well suited to to this task, enabling kilometer-scale effective cavity lengths with low loss in a millimeter scale crystal. Both the amount of phase noise and sensitivity to current noise are greatly reduced
due to the steep dispersion of the hole.

We now provide a theoretical treatment of the effect of optical feedback on the linewidth of a diode laser. We perform a similar analysis as Agrawal \cite{agra84} and consider the case where output
light from the laser is directed through an external optical network and re-injected into the diode. The intracavity laser field $E(t)$ then obeys the equation of motion
\begin{equation}
\dot{E}(t)=-i\delta E (t) + \frac{1}{2} (G-\gamma)(1-i\eta) E (t) + H(t) \ast E(t),
\label{equ:intracav}
\end{equation}
where $\delta=\omega-\omega_0$ is the relative frequency of the laser $\omega$ to the center of the spectral hole $\omega_0$, $G$ is the total rate of gain of the cavity, $\gamma$ is the cavity decay rate
and $\eta$ is the linewidth enhancement factor which is proportional to the carrier induced refractive index change. The last term in Eq.~\ref{equ:intracav} is due to the optical feedback, where $H(t)$ is
the impulse response of the external optical network and $\ast$ represents convolution.

In our case the external optical network consists of a propagation delay and a spectral hole hence $H(t)$ is given by
\begin{equation}
H(t) = \mathscr{F}^{-1} \left[ \kappa \exp\left(i[ -\omega \tau + \phi_{0}] \right) h(\omega)_{\text{h}} \right] ,
\end{equation}
where $\tau$ is the round trip time for the external path in the absence of the spectral hole, $\kappa$ is the amount of feedback light to the diode, and $\phi_{0}$ is the phase shift on the light resonant
($\omega=\omega_0$) with the hole. The transfer function due to the spectral hole is given by \cite{bottger}
\begin{equation}
h(\omega)_{\mathrm h} = \exp \left[ - \frac{\alpha L}{2} \left(1 - \frac{\Gamma}{\Gamma - i \delta }\right) \right] ,
\label{eq:ctf}
\end{equation}
where $\alpha (\omega) L$ is the optical depth and $\Gamma/2\pi$ is the width of the spectral hole. 

For the case where the laser is tightly locked near the center of the hole ($\omega\approx\omega_0$) we can replace the hole transfer function by one that is constant in amplitude and having a linear ramp
in phase. This is equivalent to replacing the external optical network by an effective coupling constant $\kappa_{\mathrm{eff}}$, delay $\tau_{\mathrm{eff}}$, and on resonance phase shift $\phi_{0\,\mathrm{eff}}$ in the time domain. In this analysis we have neglected the effect of the diffraction grating as over the narrow spectral region defined by the spectral hole its spectral response is
close to being flat. The impulse response of the external cavity with the spectral hole is then given by
\begin{equation}
H(t) \approx \mathscr{F}^{-1} \left[\kappa_{\text{eff}}\exp\left(i[ -\omega \tau_{\text{eff}} + \phi_{0\text{eff}}] \right)\right]
\label{effparam}
\end{equation}
where $\kappa_{\text{eff}}=\kappa\,\textrm{Re} \left[(h(\omega \approx  \omega_0)_{\text{h}})\right]$, $\tau_{\text{eff}}=\tau +\frac{\alpha L}{2 \Gamma}$ and $\phi_{0\,\text{eff}}=\omega_0 \tau_{\text{eff}} +\phi_{\text{m}}$. The dispersive element can therefore be characterized simply as a large cavity so $H(t)\ast E(t) \approx \kappa_{\text{eff}}E(t-\tau_{\text{eff}})e^{i \phi_{0\text{eff}}}$.

In our setup, we have used a 0.1\% Tm:YAG sample. We estimate values for the effective parameters for our experiment to be $\Gamma = 50$~kHz and $\alpha L=0.5$. Hence we get a large effective time delay $\tau_{\text{eff}}= 5\mu s$ due to dispersion of the hole. This large effective time delay is equivalent to an external cavity that is 1.5 km long. In praseodymium systems \cite{morgan}, features with tens of
kilohertz resolution have been prepared in samples with $\alpha L \approx 30$ which would lead to millisecond scale effective delays.

We can obtain the following expression for the linewidth of our laser \cite{agra84}
\begin{equation}
\Delta f=\Delta f_0 \frac{1}{\left[1+X\cos(\phi_{0\text{eff}}+\phi_R)\right]^2},\label{narrowline}
\end{equation}
where $X= \kappa_{\text{eff}}\tau_{\text{eff}} \sqrt{1+\eta^2}$, $\phi_R = \tan^{-1}\eta$, $\Delta f_0$ is the linewidth of the bare diode and $\Delta f$ is the linewidth of the laser with optical feedback. It can be seen qualitatively that for a large dispersion, $\tau_{\text{eff}} \gg 1$, and the appropriate feedback phase, $\phi_{0\text{eff}}+\phi_R \approx 2\pi m$, we can obtain a large reduction in laser linewidth.

As well as reducing the fundamental linewidth, the optical feedback also greatly reduces the sensitivity of the laser to current noise. This sensitivity is characterized by ${\partial   \omega}/{\partial \Omega}$, the change in laser frequency with the change in the resonant frequency of the diode $\Omega$. Optical feedback reduces this from one to \cite{isky}
\begin{equation}
\frac{\partial \omega}{\partial \Omega} = \frac{1}{1+X\cos(\phi_{0\text{eff}}+ \phi_R)}
\label{fmnoise}.
\end{equation}
%
\begin{figure}[h]
\begin{center}
\includegraphics[width=8.3cm]{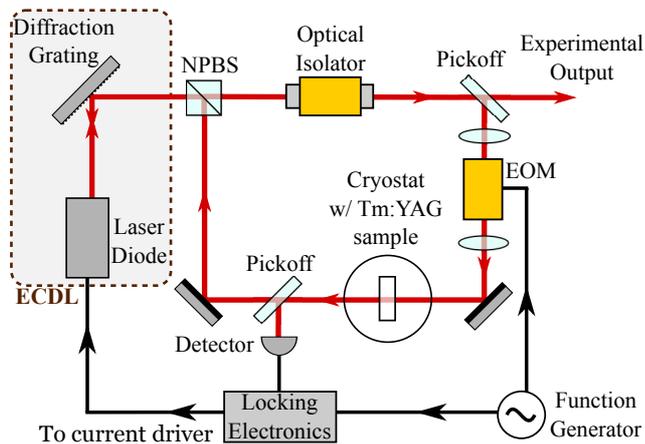}
\end{center}
\caption{(Color online) Laser locking setup. We use an extended cavity diode laser (indicated by the greyed dotted box) which is then locked to a spectral hole reference. Standard PDH\cite{pdh} locking was used, along with optical feedback.}
\label{fig:schematic}
\end{figure}
We now report on experimental results showing this reduction in linewidth. The experimental setup is shown in Fig.~\ref{fig:schematic}. We used an 80 mW single mode laser diode (Eagleyard EYP-RWE-0840-06010-1500-SOT02-0000) driven using a home-built version of Libbrecht and Hall's current supply \cite{libbhall}. The laser was built using a diffraction grating (1800 lines/mm) in a Littrow configuration and is a modified version of the system given in \cite{arnold}. We tune the laser using the grating angle, as well as tuning the temperature and current, to operate at 792 nm. The ECDL is placed within a sealed aluminum box to achieve environmental isolation.

We lock the laser to a spectral hole using PDH locking \cite{bottger,julsgaard}. To generate the error signal, a portion of the output beam is directed through an electro-optic phase modulator (EOM) to generate the frequency modulated sidebands at 30 MHz. The dithered beam is then directed through a Tm:YAG~0.1\% sample cryogenically cooled to 2.7 K, a spectral hole in which acts as the frequency discriminator. The sample used has dimensions 8$\times$4$\times$4 mm, with light propagating 4~mm along the $[1\,\bar1\,0]$ direction. The incident beam on the sample had a power of 1 mW and a diameter of 4 mm. The laser frequency is then kept at the center of the hole by using a servo loop on the laser current.  As the holes are transient, the error signal frequency response drops off at low frequencies as well as at high frequencies. Due to this we rolled the servo gain off at low ($\sim 1$ Hz) frequencies as well as at high frequencies. This ensured that the laser servo would not be driven to its rails by integrator-like transfer functions acting on small offsets.

\begin{figure}[h]
\begin{center}
\includegraphics[width=8.3cm]{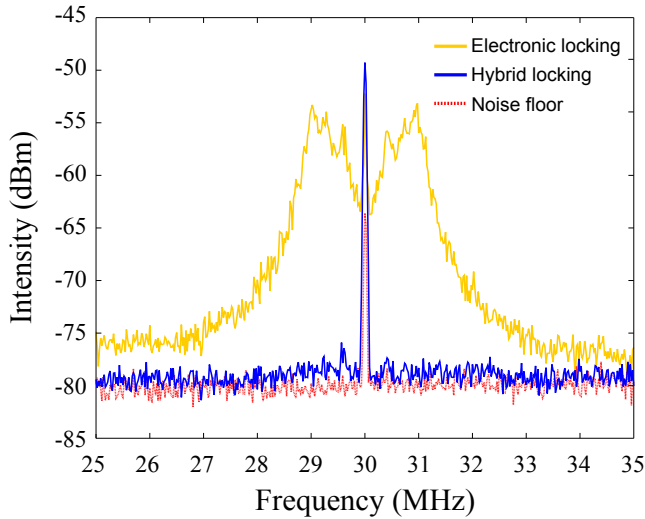}
\end{center}
\caption{(Color online) Spectrum of the signal on the locking detector. This is mixed down 30MHz to provide the locking error signal. The curves indicate: (i) optical feedback disabled (yellow), (ii) optical feedback enabled (blue), and (iii) the detector noise floor (dotted
  red) which is measured with the detector blocked.}
\label{fig:noise}
\end{figure}

To demonstrate the effects of optical feedback, a portion of the light transmitted through the crystal is directed back to the diode. We can monitor changes to the laser linewidth using the spectrum of the error
signal as shown in Fig.~\ref{fig:noise}. A dramatic reduction in the high frequency components of the error signal can be seen when optical locking is present. The resulting noise spectrum is close to being
limited by the dark noise of the detector.

\begin{figure}[h]
\begin{center}
\includegraphics[width=8.3cm]{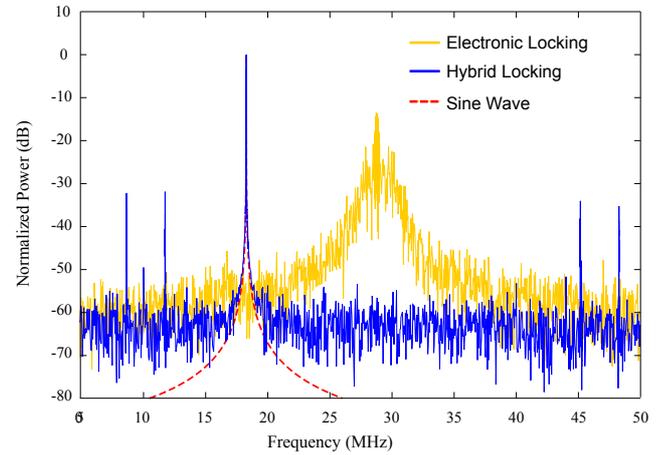}
\end{center}
\caption{(Color online) The power spectrum for two lasers, both with electronic feedback only (yellow/light) or with electronic and optical feedback applied (blue/dark).  The red dashed line is the result of applying the same windowing and FFT procedure used for the data to a sine wave.}
\label{fig:beatSpectra}
\end{figure}

To better characterize the noise on the laser, we constructed a second laser and interfered it with the first to produce a beat signal. This laser was as independent as practical, having a dither frequency of 27 MHz and using a different spot on the same crystal to prepare its spectral hole. The beat signal was amplified and frequency down-converted by mixing with a local oscillator at a frequency close
to the frequency difference between the lasers. The signal was sampled with at 100 MSamples/s, linear drift corrected for, and the power spectra, shown in Fig.~\ref{fig:beatSpectra}, then calculated.

One can see a dramatic linewidth reduction and the phase noise caused by the optical feedback (blue/dark curve), with the exception of small sidebands at $\sim10$kHz from the carrier. Indeed the spectrum of the beat signal for the optically locked lasers was either limited by the finite sampling time (1~ms) or the noise floor of -111 dBc/Hz. 

The peaks around 45 and 48 MHz are 27 and 30 MHz from the beat frequency respectively. These are due to the fact that the portion of the PDH locking sidebands are injected into the diode. The two
other peaks are aliases of the other sidebands.

Implementing the optical locking was very straight forward, apart from making the optical spot as large as possible to reduce power broadening, no matching of the optical mode to the rare earth was
needed. Furthermore no efforts to lock the phase of the large external cavity were made.  The spectral hole burning mechanism used in this work was relatively short lived, about 12 ms for the $^3$F$_4$ shelving state of Tm$^{3+}$:YAG.  Due to this and the large inhomogeneous broadening, ~30~GHz, the phase of the optical feedback did not need to be controlled. The laser and the hole simply followed the slow movements of the external cavity.

While the short term noise properties of the laser were favorable, there was significant drift ($\mathcal{O}100$~kHz/s) and occasional frequency jumps, of tens of megahertz due to mode hops of the external cavities of the lasers. However these deficiencies did not affect our subsequent applications of the laser in investigating photon echo based quantum information processing \cite{rase} and the optical detection of ultrasound \cite{tay10}. Furthermore, due to the low phase noise, only a small bandwidth servo would be needed to stabilize the laser should long term frequency stability be required.


This work was supported by the New Zealand Foundation for Research, Science and Technology under Contract No.\ NERF-UOOX0703. The authors would also like to acknowledge Sascha Hoinka for useful discussions in building the diode laser.

\end{document}